\documentclass[conference]{IEEEtran}
\IEEEoverridecommandlockouts
\usepackage{cite}
\usepackage{amsmath,amssymb,amsfonts}
\usepackage{algorithmic}
\usepackage{graphicx}
\usepackage{textcomp}
\def\BibTeX{{\rm B\kern-.05em{\sc i\kern-.025em b}\kern-.08em
    T\kern-.1667em\lower.7ex\hbox{E}\kern-.125emX}}
    
\begin{document}

\title{ Blockchain-Empowered Immutable and Reliable Delivery Service (BIRDS) Using UAV Networks}
\author{\IEEEauthorblockN{Sana Hafeez, Habib Ullah Manzoor, Lina Mohjazi, Ahmed Zoha, \\
Muhammad Ali Imran and Yao Sun\\}
\IEEEauthorblockA{James Watt School of Engineering, University of Glasgow, Glasgow, United Kingdom\\ Email: {\{s.hafeez.1, h.manzoor.1\}@research.gla.ac.uk},\\ {\{Lina.Mohjazi, Ahmed.Zoha, Muhammad.Imran, Yao.Sun\}@glasgow.ac.uk}
}
\vspace{-\baselineskip}}

\maketitle

\begin{abstract}
Exploiting unmanned aerial vehicles (UAVs) for delivery services is expected to reduce delivery time and human resource costs. However, the proximity of these UAVs to the ground can make them an ideal target for opportunistic criminals. Consequently, UAVs may be hacked, diverted from their destinations, or used for malicious purposes. Furthermore, as a decentralized (peer-to-peer) technology, the blockchain has immense potential to enable secure, decentralized, and cooperative communication among UAVs. With this goal in mind, we propose the Blockchain-Empowered, Immutable, and Reliable Delivery Service (BIRDS) framework to address data security challenges. BIRDS deploys communication hubs across a scalable network. Following the registration phase of BIRDS, UAV node selection is carried out based on a specific consensus proof-of-competence (PoC), where UAVs are evaluated solely on their credibility. The chosen finalist is awarded a certificate for the BIRDS global order fulfillment system. The simulation results demonstrate that BIRDS requires fewer UAVs compared to conventional solutions, resulting in reduced costs and emissions. The proposed BIRDS framework caters to the requirements of numerous users while necessitating less network traffic and consuming low energy.
\end{abstract}


\section{Introduction}
\IEEEPARstart {T}{hanks} to efficient and high mobility, unmanned aerial vehicles (UAVs) are taking on essential tasks, including search and rescue, remote sensing, and delivery\cite{unknown}. 
Anticipating a logistical challenge in achieving rapid and cost-effective delivery, commercial enterprises are increasingly exploring drone technology to reduce delivery times and costs. UAV deployment proves efficient for final-mile delivery, considering environmental and economic factors, making it a promising aerial solution \cite{Moadab2022}. 
Traditional delivery vehicles have become infeasible due to high fuel costs, city troubles, and environmental consequences due to urban distribution problems \cite{dai2022joint}. Recently, UAVs are poised to play a pivotal role in achieving efficient and swift delivery services, leveraging their distinct attributes of high mobility, adaptable deployment, and cost-effectiveness. Their mobility also positions UAVs to serve as airborne communication platforms, enhancing connectivity for ground-based operations.

Nevertheless, UAVs can serve as intermediaries connecting ground users and spacecraft through Line-of-Sight (LoS) channels, suggesting a pivotal role for communication-based UAVs in 6G networks. The substantial power consumption challenge in wireless networks, particularly for long-distance transmissions with notable path loss, is acknowledged. The constrained power capacity of small UAVs has further propelled recent research interest in the realm of environmentally conscious UAV communication \cite{hafeez2023bcsfl}.
 Despite recent discussions on solutions, [4]-[9], privacy and security aspects remain incomplete. A significant challenge in UAV-enabled delivery services is the potential exploitation by malicious actors through hacking and malware \cite{9860313}.
 A promising solution to address these issues is the involvement of blockchain. As a distributed decentralized network, the blockchain provides user privacy, immutability, transparency, and reliability \cite{10182294}. Moreover, blockchain incorporates many desirable properties of a back-end solution at once, including decentralization, redundancy, fault tolerance, security, and scalability. As in the research area of UAVs, the application of conventional blockchain exposes a significant problem, since multiple nodes are already resource-constrained UAVs. 
  In \cite{hafeez2022beta}, we introduce a BETA-UAV blockchain-based efficient authentication for secure UAV communication. The objective is to enable mutual authentication and freshness identification so that the UAV network can establish secure communication channels. Proof-of-freshness (PoF) or authentication protocols allow UAVs to integrate with these systems with minimal hassle and maximum security.
 
 Following this, the authors in \cite{yu2023laka} create a robust and lightweight authentication and key agreement scheme for the cloud-assisted UAV using blockchain in a flying ad-hoc network (FANET) to guarantee data sharing decentralization and integrity. By streamlining these challenges, the proposed BIRDS framework can be used as an innovative solution to overcome them.  
The BIRDS framework uniquely integrates UAV registration, blockchain-based authentication, node selection, and reputation scoring for heightened scalability, energy efficiency, and security. This distinguishes it from the state-of-the-art, as confirmed by rigorous simulations, by offering an innovative solution that combines privacy, security, and UAV operational efficiency. The BIRDS framework provides end-to-end security spanning rigorous authentication, optimized node assignment, and performance-based reputation management to enable safe, dependable, and energy-efficient delivery operations.
Our main contributions to this paper are summarized below.
 \begin{itemize}
     \item We develop the BIRDS framework, which endures
a detailed verification and registration, benchmark comparison, and eventual proof-of-competence (PoC) generation, and provides solutions to traditional overheads.
     \item We introduce PoC as an advanced and credible consensus mechanism, ensuring both scalability and energy efficiency.
     \item BIRDS consists of four stages: Initiating with secure UAV registration, it proceeds to blockchain consensus inspection, ensuring dual security for both UAVs and user-side registration. Subsequently, it involves UAV node selection and culminates in awarding a reputation score.
     \item We examine the novelty of our approach through competency and reputation scores. UAV node selection relies on key parameters: Timestamp, proof-of-identification (PoI), proof-of-resources (PoR), and delivery time.
     \item Finally, we perform numerical simulations to evaluate the performance of BIRDS compared with classical blockchain consensus algorithms and demonstrate its supremacy in delivery time and energy consumption. 
 \end{itemize}

\begin{figure*}[t]
    \centering    \includegraphics[width=0.93\textwidth]{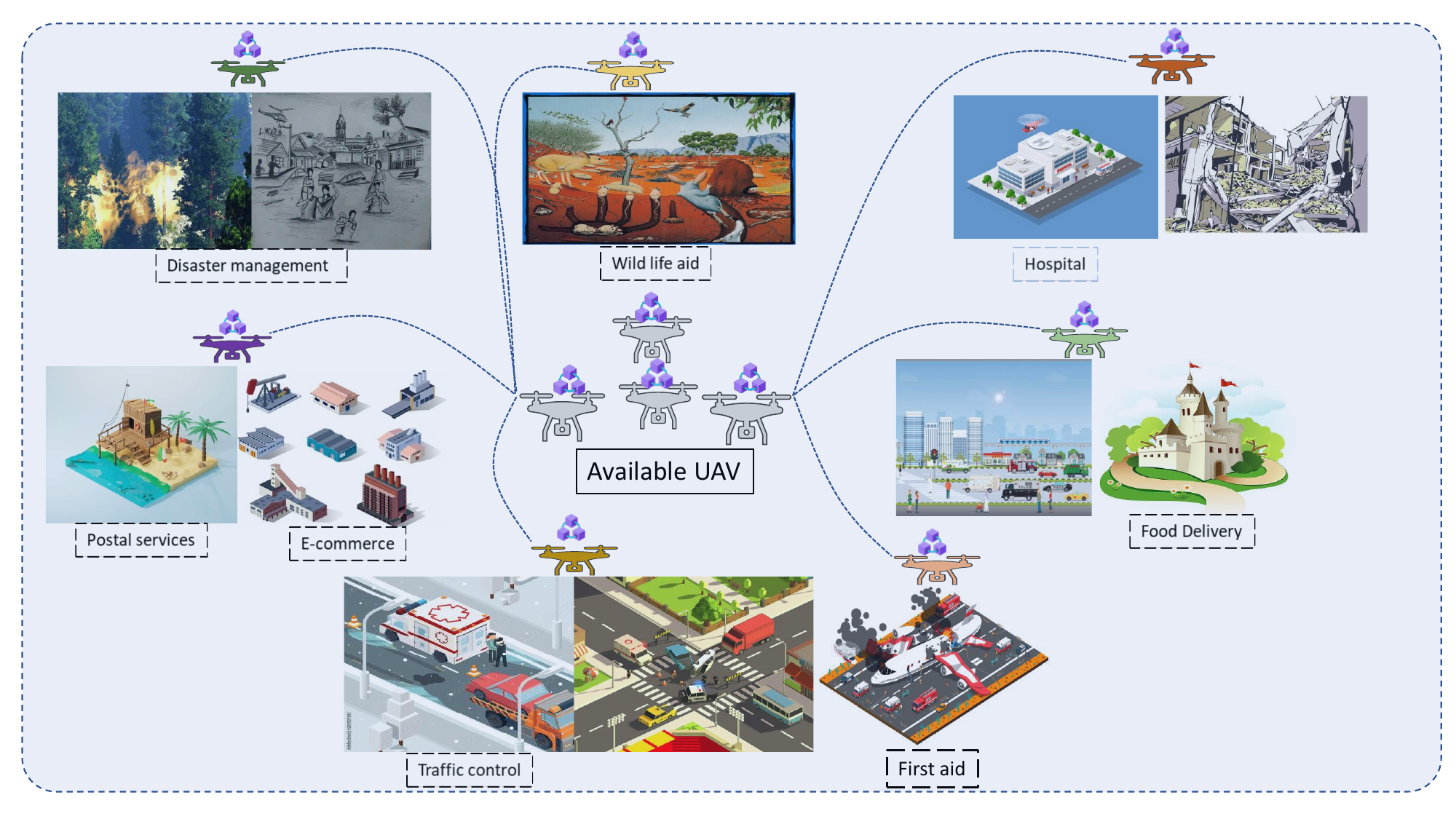}
    \caption{Blockchain-Assisted UAV Delivery Services.}
    \label{fig:data}
\end{figure*}
\begin{table}
    \centering
    \small
    \caption{List of Notations.}
    \begin{tabular}{|c|c|}
    \hline \text{ Notations } & \text{ Descriptions } \\
    \hline 
    $\mathcal{U}$ & UAV array \\
    $\mathbb{I}$ & Set of users \\
    $\mathcal{D}$ & Group of delivery data \\
    $\mathcal{Q}$ & Channel resources set \\
    $\mathbf{k}_v$ & Average riding speed/velocity of UAV v \\
    $d$, $q$ & Data packet, Coverage  \\ 
    $\mathfrak{f}_u^d$ & Flying distance between $d$ and $u$ \\
    $\mathbf{E}_u^{\text{R}}$ & UAVs leftover energy of $R$ and $u$ \\
    $p_i^t$ & Transmission power among $i$ and $t$\\
    $h_i^q$ & Channel power gain between $i$ and $q$\\
    $L_u^i$ & Link among UAV $u$ and user $i$\\
    $\gamma$ & Additive white Gaussian noise \\
    $\mathcal{T}_i^u$ & User's achievable transmission rate\\
    $\mathbb{B}$ & Channel Bandwidth\\
    $\mathfrak{f}_U^e$ & Flying usage of UAV between $e$ and $U$\\
    $H_U^e$ & UAV hovering expenditure of $e$ and $U$\\
    $\gamma_\mathfrak{f}^p$ & Cost of energy per unit distance traveled\\
    $\gamma^h$ & Hovering energy used per unit time\\
    $W(r), i(r)$ & Weighted reward and each user reward\\
    $e_u^{\text{T}}$ & Total energy consumption of UAV\\
    $\mathbf{E}_u^{\max}$ & Maximum potential energy\\
    $\tau_i^d$ & Transmission delay for delivering data\\
    $\chi_0$ & Timeline of for delivering data\\
    $\mathbb{S}_i$ & Parameter of satisfaction degree \\
    $\delta(r)$ & Reward for successful consensus\\
    $\mathcal{U}_i\left(s_i^c\right)$ & Utility of UAV for delivering data\\
    $\mathcal{D}_i\left(s_i^c\right)$ & Utility of user for delivering data \\

    \hline
    \end{tabular}
    \label{tab:my_label}
    \vspace{-5mm}
\end{table}
The rest of the paper is structured as follows.
Section II covers the system model and problem description UAV network model, communication, and mobility model.
Section III introduces the proposed BIRDS scheme, which briefly covers all BIRDS insights.
The results and discussion are explained in Section IV and the conclusions are presented in Section V.
    
\section{System Model and Problem Description}
Section II presents and explains the specifics of the system modeling and framework, covering all three aspects: the UAV network model, the communication model, and the mobility model.
\subsection{UAV Network Model}
In Fig. 1, the BIRDS delivery scenario involves adapting ground infrastructures for diverse deliveries, leading to communication complexities. Mobile network operators strategically deploy aerial base stations (ABS) and equipped UAVs, facilitating efficient communication and data delivery services to individuals within UAV networks.

\textbf{UAV Selection}: UAVs deliver communication and data services from the air when a UAV delivery service is launched. BIRDS selects a credible UAV node so that the UAV behaves according to the ground infrastructures, Let $\mathcal{U} = \{1, \ldots, u, \ldots, U \}$ denote the UAV array.

\textbf{Customers}: Users access cellular networks in emergency or disaster zones for delivery operations, and massive amounts of high-value data on wireless devices are at risk. Therefore, users require communication and data delivery services to diminish the risk of data breaches. $\mathcal{I} = \{1, \ldots, i, \ldots, I \}$ represents the set of users in the area considered.
To permit users to communicate with UAVs, we consider large-scale emergency networks or a symmetric directed graph where $N$ is the group of nodes, i.e. UAVs and users.
\subsection{Communication Model}
Every UAV creates multiple communication channels, and the series of channels within the UAV's coverage is denoted as  $\mathcal{Q}=\{1,\ldots, q,\ldots,Q\}$. Multiple users can simultaneously access the channels, and as a result, channel interference from other users occurs during data delivery. The signal-to-noise ratio (SNR) between user $i$ and UAV $u$ can be calculated as follows.
\begin{equation}
	 \zeta_i^u=\frac{p_i^t h_i^q}{\sum_{i = \in \mathbb{I}} h_i^q p_i^t+\gamma}.
\end{equation}

Consequently, the available user transmits rate on track $p$ for data delivery is decreased by $u$. The user's achievable transmission rate can be expressed as
\begin{equation}
 T_i^u = \mathbf{B} \log \left(1 + \zeta_i^u\right).  
 \label{eq7}   
\end{equation}
Data delivery in the coverage:
 The delivery of set of data packets is represented by $\mathcal{D} =\{1,\ldots, d,\ldots, D\}$. The data size transmitted by the user $i$ is denoted as $s_i^d$. $G$ represents the total amount of data delivered within the UAV's coverage area $\sum_{i\in\mathcal{I}, d \in \mathcal{D}}\left\{\mathbb{S}_i ^d\right\}$. UAV hovers in a specific spot through data delivery, and its altitude determines the effects of data delivery.
 Transmission delay in delivery of  data
UAV hovers in a fixed location during data delivery and the transmission delay for delivering data is determined by its altitude, where the transmission delay is shown by
\begin{equation}
    \chi_i^d=\frac{s_i^d}{T_i^u}.
\end{equation}
The data transmission delay $\mu$ outperforms the target time.
The target time for \(\mu\) is given by \(T^{0}\), which indicates that the data cannot be delivered if the transmission delay exceeds \(T^{0}\).
 The primary role of BIRDS is to autonomously differentiate reliable and unreliable UAV delivery channels in a decentralized manner. Specifically, the architecture employs smart contracts to facilitate the initialization and registration of UAVs or clients.
\noindent
\subsection{Mobility Model}
The mobility model of UAVs incorporates communication and data delivery services. The UAV flight path consists of multiple distinct points in a three-dimensional Cartesian coordinate system.
The velocity of a UAV is denoted by $\mathbf{k}_v=\left\{k_v^x, k_v^y, k_v^z\right\}, \forall u \in \mathcal{U}$. Here, $k_u^x, k_u^y$, and $k_u^z$ represent the specific speeds of the UAV in 3D Cartesian coordinates. A UAV's flying distance can be calculated as
\begin{equation}
\mathfrak{f}_u^d=\tau_u^f\left\|\mathbf{k}_u\right\|_2, \forall u \in \mathcal{U},
\end{equation}
where $\tau_u^f$ is the duration of the UAV's flight. Here, the energy consumed by a UAV is $e_u^{\text{T}}$, which depends on the flight power $\sigma$ as well as the hovering expenditure is presented by $\mathbb{H} = \gamma^h \tau_i^d$.
Accordingly, the overall energy consumption of UAVs that provide users with data and communication delivery services can be written as follows:
\begin{equation}
    e_u^{\text {T }} = \sigma +\gamma^h \tau_i^d .
\end{equation}
For UAVs to be able to return for recharging, their remaining energy must exceed a predetermined threshold
\begin{equation}
\mathbf{E}_u^{\text{R}} > E_u^{\mathrm{thr}}.
\end{equation}
\begin{figure*}[!t]
    \centering
 \includegraphics[width=0.95\textwidth]{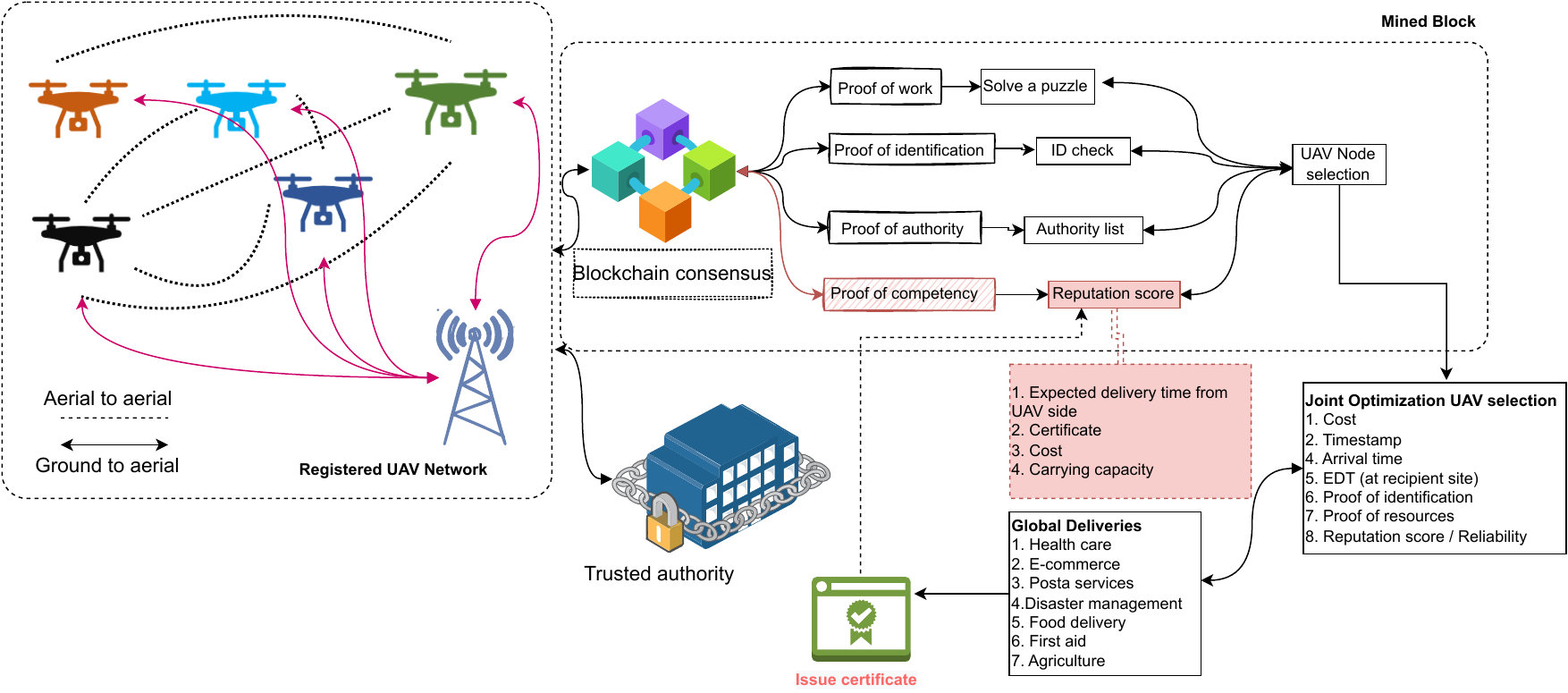}
    \caption{BIRDS Framework. }
    \label{fig:framework}
\end{figure*}
\vspace{-\baselineskip}
\section{Proposed BIRDS Scheme}
Based on the network model explained in Section II, We present BIRDS, a decentralized framework designed for identifying feasible UAV delivery routes. BIRDS comprises three primary contracts: Registration of UAVs, trusted authority (TA), and registered UAV identities. In addition, we leverage the subsequent blockchain consensus stage, along with an automated delivery tracker or UAV reputation scores, to execute the third BIRDS phase, involving UAV node selection for optimal job assignment.
\subsection{Authentication and Registration Phase of BIRDS}
The initial stage of the BIRDS framework involves UAV authentication and registration. In a network comprising 20 UAVs, each UAV is registered with details including node ID, size, weight, battery capacity, flight duration, and travel distance. BIRDS supports both aerial-to-aerial (A2A) communication and aerial-to-ground (A2G) delivery, as presented in Fig. 2. The red lines represent A2A communication links between the UAVs, showing how they can communicate with each other during flight. Different colors assigned to UAVs signify their states: green denotes readiness, blue signifies information transmission, orange indicates queueing, and black represents registration entry and waiting. Once registration is complete, participants proceed to the subsequent phase, focusing on security and privacy, which are fundamental attributes of BIRDS.

\subsection{Proof-of-Competence in BIRDS}
We provide greater transparency, immutability, and communication efficiency by comparing three consensuses; a detailed discussion is given in \cite{khan2021systematic}. In node selections, proof-of-work (PoW), proof-of-identity (PoID), and proof-of-authority (PoA) are compared.
Concluding the process, we present the PoC consensus, a novel solution that encompasses the requisite competencies to tackle blockchain challenges. Based on specific contract conditions, UAV nodes meeting these criteria are selected based on the merits of BIRDS, qualifying for the intermediate stage leading to the security assessment.

\subsection{BIRDS Criteria for Miners}
The key aspects of the BIRDS blockchain design are as follows:
The BIRDS framework utilizes a customized blockchain structure to enable secure and reliable UAV delivery services. Several architectural choices in the BIRDS blockchain design aim to enhance security, efficiency, and reliability.
First, the block identification mechanism uniquely labels each block to avoid ambiguity and enable tracking. The BIRDS $ID$ assigns a distinct identifier to every block added to the chain.
Second, BIRDS leverages cryptographic hashing techniques to guarantee the integrity of blockchain data. A Merkle tree root hash securely consolidates all transaction hashes into one final checksum. This preserves validity across blockchain updates.
Third, chaining hash pointers chronologically link blocks through one-way cryptographic hashes. Each new block contains the hash of the previous one, creating an immutable record of events. The chaining of hashes also allows efficient data lookups.
Furthermore, accurate time-stamping of blocks through BIRDS timestamps maintains an auditable timeline of blockchain activity. Timestamps impart sequence to block generation.

In addition, BIRDS allows variable transaction sizes through customized block headers and transaction lists. This provides flexibility and efficiency in block creation and chaining.

Finally, the consensus mechanism in BIRDS is designed for scalability through tweaked difficulty levels and hash requirements. This is critical for enabling low-latency confirmation of blocks.

\subsection{Credibility of UAV Node Selection}

In the BIRDS framework, after populating the blockchain with deanonymized UAV data and encrypting device data, an attacker leverages machine learning on the blockchain. To enhance privacy, we propose techniques like multi-node ledgers, transaction delays, and obfuscation using cryptographic methods such as blind and ring signatures. UAV communication transactions are initiated by nodes, which can be data-empty or encrypted, ensuring adherence to communication protocols. The nodes, which can be data-empty or encrypted and ensure adherence to communication protocols, start UAV communication transactions. Registered UAVs must validate their prior identity for a new one, addressing challenges. The selected UAV gains reputation scores, prioritizing reputable UAVs for subsequent task allocation.

\subsection{Energy Consumption in BIRDS}
  
BIRDS aims to improve the sustainability of blockchain systems by mitigating energy-intensive miner operations, a primary contributor to overall consumption. Energy usage, quantified by the power of the framework, impacts work efficiency. The energy consumption of individual miners can be calculated using the following formula based on this concept.
\begin{equation}
     E\eta =  \frac{E_{u}} {E_T},
\end{equation}
 \begin{equation}
     E_T =\frac{e_u^{\text {T }}}{E_{max} - E_u},
\end{equation}   
\begin{equation}
e_u^{\text{T}} = \frac{P^T} {E_\mu^{\text{max}} - p_i^t}.
\end{equation}
where $e_u^T$ represents the total energy consumption of the UAV and $\mathbf{E}_u^{\text{max}}$ is the maximum potential energy. \\
The reward function in BIRDS is designed to reduce system-weighted cost and optimize device energy provision, helping to make better judgments. In this work, we specify the instant rewards as
\begin{equation}
    I(r)= \begin{cases} \delta (r)-k \cdot \frac{W(r)}{i(r)}-\rho(r), & \text{if } UAV_n(r) \leq t_l \\ -k \cdot \frac{W(r)}{i(r)}-\rho(r), & \text{if } UAV_n(r)>t_l \end{cases},
\end{equation}

where $\delta (r)$ represents the rewards for a successful consensus if such consensus is shorter than the time limit. Moreover, $k$ is the weighted coefficient of the system cost. Finally, $\rho(r)$ denotes the penalty rewards and is given as
\begin{equation}
  \rho(r)=p \frac{\mathbf{E}_u^{\text{{avg}}} - \mathbf{E}_u^{\text{{R}}}}{\gamma^f} ,
\end{equation}
where $p$ is the penalty index that defines the ratio of rewards in the penalty $n$ reward function, $\mathbf{E}_u^{\text{avg}}$ represents the average energy of all devices, and $\gamma^f$ represents the energy consumed per unit of time while hovering.
  \vspace{-\baselineskip}\begin{figure}[!ht] 
         \centering
         \includegraphics[width=0.45\textwidth]{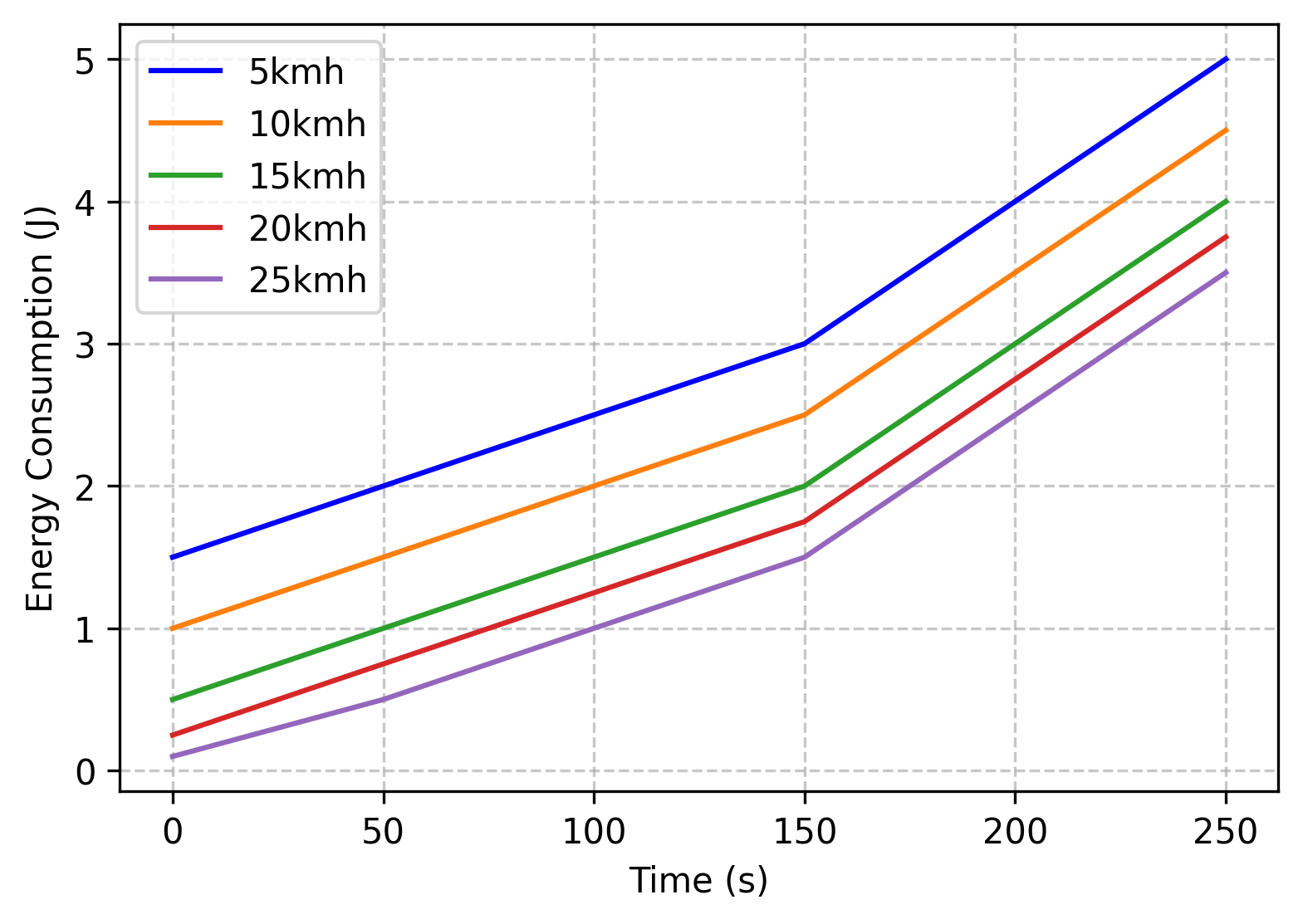}
	 \caption{Energy Consumption vs Time.}
 	\label{delay}
\end{figure}
\subsection{UAV Reputation Score in BIRDS}
Within the BIRDS communication framework, consistent data transmission is essential for block commitment, require resource allocation for block adoption. Delays in block commitment can arise from technical factors such as constrained bandwidth, computational resources, restricted throughput, and public blockchain latency.
\begin{equation}
    Rep_i =EDT+Cv+cost+K,
\end{equation}
In this context, we define $Cv$ as the certificate value and $K$ as the carrying capacity. Furthermore, consensus among UAV nodes regarding the ledger state is imperative. Thus, BIRDS solutions have been introduced to tackle these complexities.

Our model ensures a dynamic reputation set for reliable UAVs. The BIRDS reputation score is influenced by factors such as the actual delivery time ($ADT$), the delivery cost, the carrying capacity, and $Cv$. Initial scores of zero are assigned to unverified sources, with the ability to change over time. In a registered drone network, drones are categorized by mission (e-commerce, emergency communication, delivery services, healthcare, etc.) based on attributes like size, cost, and power consumption. The selected drone overcomes obstructions and receives a reputation score and certificate, contributing to our reputable BIRDS framework's UAV pool.

\section{Results and Discussion}

The simulation was conducted using MATLAB, where we modeled the behavior of 20 UAVs with diverse capabilities, considering factors like payload, velocity, and flight duration. We analyzed 20 UAVs of diverse capabilities, assuming universal low-altitude takeoff and landing. Each UAV exhibits a capacity range spanning from 1 kg to 15 kg, achieved at velocities between 400 mph and 100 mph across different payload capacities. In particular, a fully charged drone operating at maximum payload can sustain flight for a duration of one hour. In a 100 km$^2$ region, we considered 80 randomly spaced waypoints. The evaluation of the effectiveness and design of our strategy is based on three established metrics: energy efficiency, reputation score, and scalability.

 \begin{figure}[!ht] 
        \centering
        \includegraphics[width=0.45\textwidth]{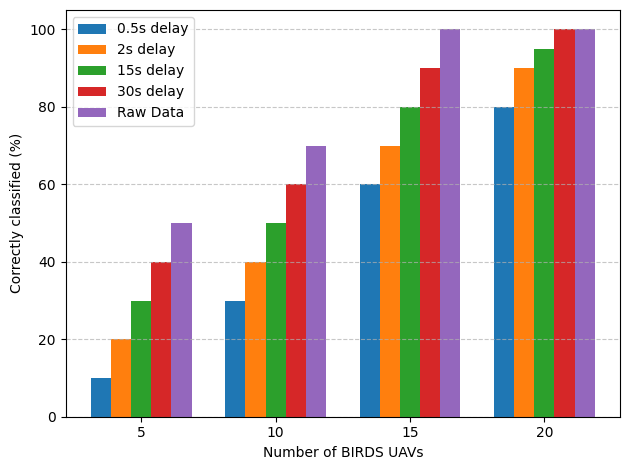}
	\caption{Impact of the Number of UAVs on the Delay.}
	\label{delay}
\end{figure}

\begin{figure}
    \includegraphics[width=0.45\textwidth]{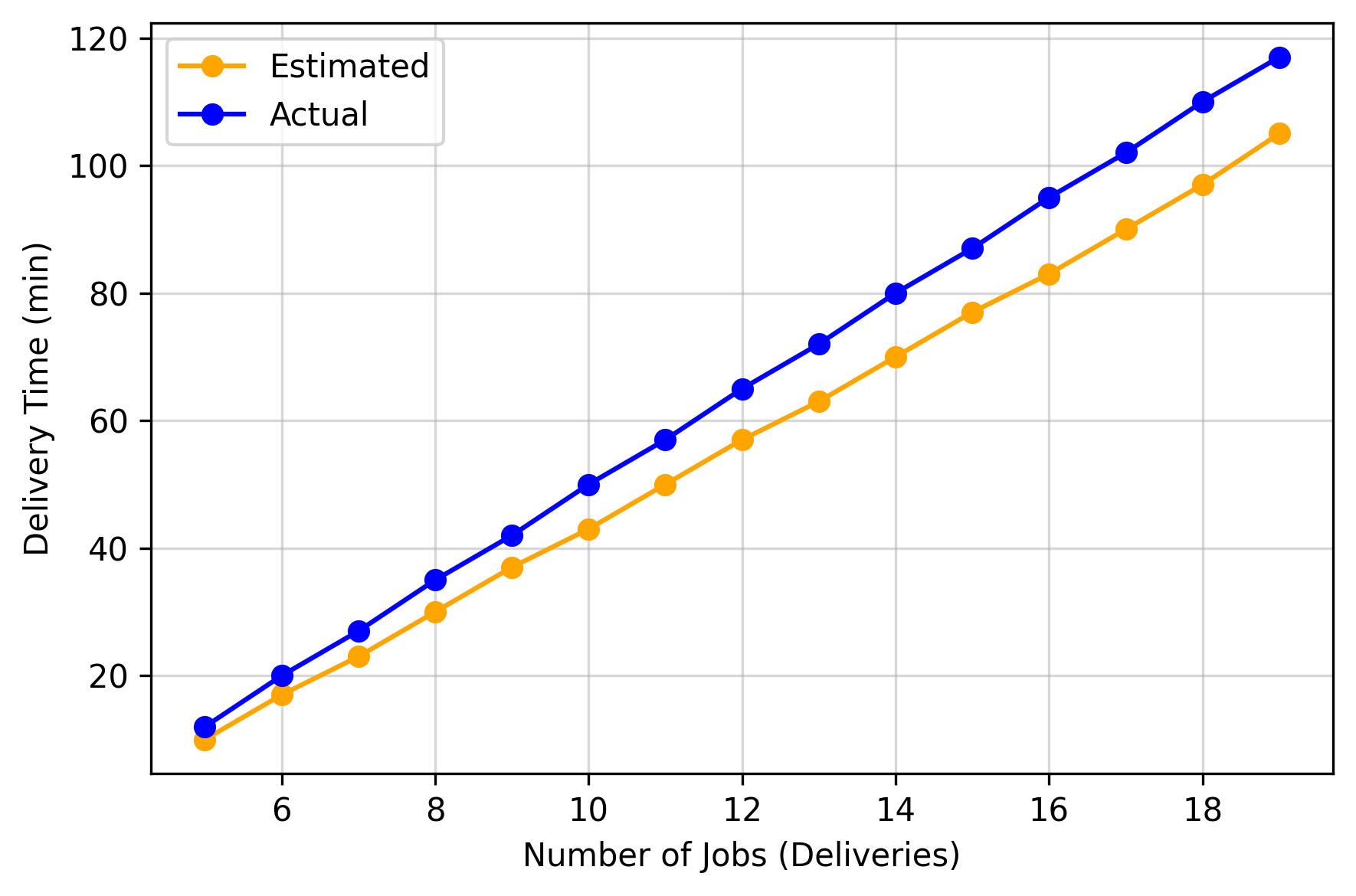}
   \caption{Estimated and Actual Delivery Time for UAV Tasks.}
   \label{deliverytime}
\end{figure}

 \begin{figure}[!t]
      \includegraphics[width=0.45\textwidth]{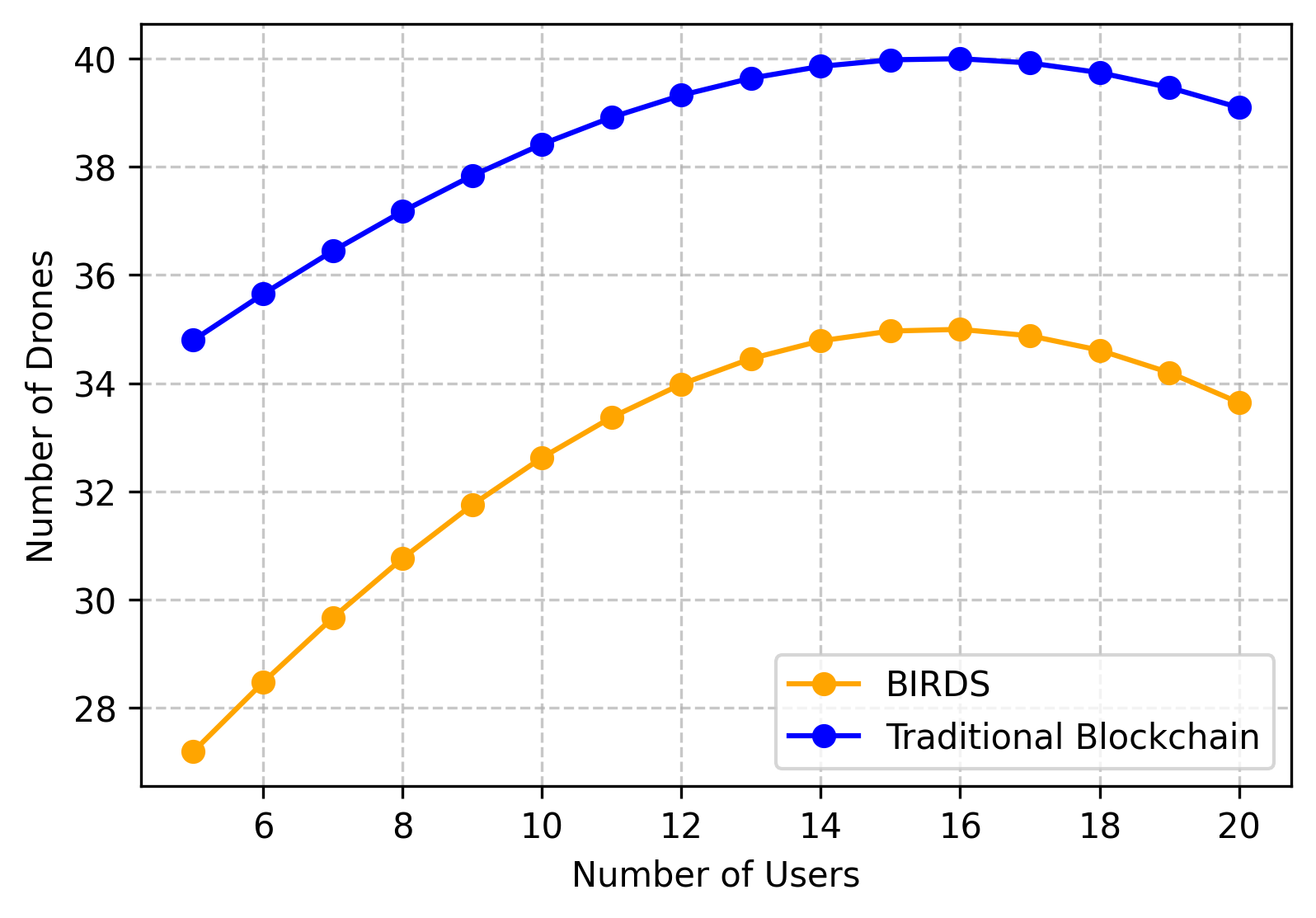}
 	\caption{Traditional Blockchain vs BIRDS.}	\label{bcconcensus}

\end{figure}
The variation in UAV node count per ledger (x-axis) demonstrates significant accuracy in initially categorizing device speeds, especially with two or four devices. The success rate exceeds 50\% when employing more than 15 UAVs. As depicted in Fig. \ref{delay}, this matrix effectively reduces the associated probability. Subsequently, an intricate multipart transaction strategy is employed to detect deployed devices and mitigate risks using multiple packets. This approach enhances the ledger's adaptability to various user profiles, thereby augmenting the transactional dynamics. Fig. 5 displays the job count (deliveries) on the x-axis and delivery time on the y-axis. Efficient initial job processing gives way to a bottleneck beyond 15 jobs. The effectiveness of the PoC across various consensus mechanisms addresses these challenges. Fig. 6 presents an efficiency comparison between the BIRDS framework and traditional blockchains (PoW, PoID, PoA) as user counts increase. Remarkably, the BIRDS network requires fewer UAVs under heightened user loads, highlighting its ability to manage multiple users while effectively mitigating network congestion.

\section{Conclusions Future Work}
This study introduces the BIRDS framework, an innovative lightweight blockchain solution for enabling secure and reliable UAV delivery services. Through rigorous analysis, BIRDS demonstrates resilience against security threats and underscores the effectiveness of the novel PoC algorithm in ensuring scalability and energy efficiency.
BIRDS effectively addresses the energy consumption concerns associated with UAV delivery operations by optimizing throughput for rapid delivery, cost reduction, and environmental sustainability. A comprehensive performance evaluation assessing the job delivery capabilities across a diverse set of UAVs is conducted. The initial classification success rate with respect to device speed exhibits remarkable proficiency, especially in dual or quadruple device scenarios, while gradually stabilizing at 50\% as the UAV count exceeds 15.
Additionally, compared to existing schemes, the presented framework reduces communication costs, ensures lightweight computation and storage overheads, and provides superior security attributes. Moreover, it enables secure transactions between clients and UAVs for deliveries from both communication and blockchain perspectives.
While BIRDS demonstrates promising performance, certain limitations present avenues for future work. First, evaluating BIRDS in real-world large-scale UAV delivery scenarios with hardware implementations would lend further credibility. Second, enhancing the machine learning capabilities for UAV profiling and job assignment could improve overall workflow automation. Third, accommodating diverse aerial vehicle types through adaptive protocols and algorithms can augment the framework's versatility. Fourth, integrating edge computing solutions could help address latency and bandwidth constraints. Finally, advancing the reputation system with more parameters and adaptive weighting schemes can enrich trust management across decentralized UAV platforms.


\vspace{12pt}

\end{document}